\documentclass[12pt]{iopart}
\usepackage{bm}
\usepackage{iopams}
\usepackage{graphicx}
\usepackage{subfigure}
\usepackage{units}

\begin{document}
\title[Radiation force of Bessel beams]{Off-axial acoustic radiation force of pressor and tractor Bessel beams on a sphere}
\author{Glauber T. Silva$^{1}$, J. Henrique Lopes$^1$,  Tiago P. Lobo$^1$, and Farid G. Mitri$^2$}
\address{$^1$Physical Acoustics Group, Instituto de F\'isica, Universidade Federal de Alagoas, 
Macei\'o, AL, Brasil 57072-970.\\
$^2$Los Alamos National Laboratory, 
Materials Physics and Applications Division, 
MPA-11, Sensors \& Electrochemical Devices, 
Acoustics \& Sensors Technology Team, MS D429, 
Los Alamos, NM, USA 87545.}
\ead{glauber@pq.cnpq.br}

\begin{abstract}
Acoustic Bessel beams are known to produce an
axial radiation force on a sphere centered on the 
beam axis (on-axial configuration) that exhibits both ``pressor'' and ``tractor'' behaviors.
The pressor and the tractor forces are oriented along the beam's direction
of propagation and opposite to it, respectively.
The behavior of the acoustic radiation force generated by Bessel beams when the sphere lies 
outside the beam's axis (off-axial configuration) is unknown.
Using the 3D radiation force formulas given in
terms of the partial wave expansion coefficients 
for the incident and scattered waves, both axial
and transverse components of the force exerted on a silicone-oil sphere are obtained for
a zero- and a first-order Bessel vortex beam.
As the sphere departs from the beam's axis, the tractor force becomes weaker.
Moreover, the behavior of the transverse radiation force
field may vary with the sphere's size factor
$ka$ (where $k$ is the wavenumber and $a$ is the sphere radius). 
Both stable and unstable equilibrium 
regions around the beam's axis are found depending on 
$ka$ values.
These results are particularly important for the design of acoustical tractor beam devices operating with Bessel beams.
\end{abstract}

\maketitle

\section{Introduction}
The transfer of momentum from electromagnetic~\cite{klima395} and acoustical~\cite{torr402} propagating waves upon 
reflection or absorption from a particle induces
forces and torques that may be used under various circumstances to accelerate~\cite{ashkin156}, 
rotate~\cite{parkin525,mohanty563, mitri:026602, silva:epl}, trap~\cite{chu685,ashkin4853}, 
levitate~\cite{ashkin586} or even stretch~\cite{guck767} the particle itself. 
The idea of attracting/repelling objects from a distance with radiation (known as tractor/pressor beams) 
has also captured the interest of many scientists and news media~\cite{wilk12,block493,sukhov3847,corporation}. 
The pressor behavior is somehow intuitive, such that the particle facing the incident beam is expected to
be pushed away from the source. 
Alternatively, the tractor behavior occurs as a consequence of a negative force that pulls the particle toward the source.

In acoustics, various investigations revealed effects similar to both pressor and tractor beams. 
For the pressor behavior, a large body of experimental and theoretical work has been already
published (see Ref.~\cite{doinikov}).
On the other hand, it was demonstrated that a curved-wave front, such as a progressive spherical diverging wave 
incident upon a rigid or solid sphere, induced a negative (attractive or tractor) axial radiation force depending on both 
the frequency of the incident field and the distance from the source~\cite{embleton40,embleton46,hasegawa937,chen:713}. 
The same has been demonstrated for rigid cylindrical structures in a progressive cylindrical diverging 
waves~\cite{mitri620,mitri523}. 
A plane progressive wave can induce an axial negative force on an elastic sphere coated with a 
viscoelastic (absorptive) layer~\cite{mitri379,mitri337}. 
Also, the presence of a boundary close to a spherical shell has shown to induce a negative axial acoustic 
radiation force that is affected by the frequency of the incident plane waves, the distance to the boundary and 
its corresponding reflection coefficient~\cite{miri:301}.
In addition, when the incident beam is in the form of a zeroth-order (non-vortex) Bessel 
beam, the axial force can pull fluid spheres toward the source~\cite{marston3518}.
Moreover, the axial negative force of higher-order Bessel (vortex) beams on rigid~\cite{mitri1059}, 
fluid and elastic spheres~\cite{mitri245202} occurs depending on the beam's half-cone angle $\beta$
and the sphere's size factor $ka$. 
An attractive  radiation force can be also 
achieved by using highly focused acoustical tweezers~\cite{lee073701}.

In this work, the 3D radiation force caused by a zero- and a first-order Bessel beams upon 
an absorbing fluid sphere (silicone) is computed.
In doing so, the analytical expressions for the axial and transverse components of the radiation 
force~\cite{silva:3541} are used. 
These expressions are given in terms of the beam-shape coefficients of the incident wave and the 
scattering coefficients of the sphere. 
The beam-shape coefficients are evaluated through the discrete spherical harmonic transform~\cite{colton:p79} developed 
in the context of the arbitrary acoustic scattering theory of Bessel beams by a sphere~\cite{silva:298,mitri392}. 
The computations of the radiation force components for
the silicone-oil sphere reveal that the attractive force of the tractor beam decays as the sphere departs from the beam's axis.
In addition, the transverse radiation force field
may exhibit stable and unstable equilibrium regions
in the vicinity of the beam's axis.
This study may potentially assist in designing specific acoustical probes operating with Bessel beams for biological 
imaging and particle manipulation applications.
 

\section{Acoustic radiation force}
Consider an incident acoustic wave of angular frequency $\omega$ scattered by 
a fluid sphere of radius $a$ suspended in the 
wavepath within an ideal fluid.
The pressure of the incident wave is given by $p_i=\hat{p}_i e^{-i \omega t}$, where $\hat{p}_i$ is the pressure amplitude, $i$ is the imaginary unit, and
$t$ is the time. 
In turn, the fluid is characterized by the ambient density $\rho_0$ and the adiabatic speed of sound $c_0$
The 3D radiation force generated on a suspended sphere can be written in the form ${\bf f} = \pi a^2 E_0 ( {Y}_x,{Y}_y,{Y}_z)$,
where $E_0$ is the characteristic energy density of the wave, $Y_x$, $Y_y$ and $Y_z$ are the radiation force functions
in Cartesian coordinates.
These functions are given by~\cite{silva:3541}
\begin{eqnarray}
Y_x = \textrm{Re}\{Y_\perp\}, \\
Y_y = \textrm{Im}\{Y_\perp\}, \\
Y_z = \frac{1}{ \pi (ka)^2}  \textrm{Im} \sum_{l,m} (a_l^m+s_l^m) \left(  s_{l+1}^{m*}c_{l+1}^m - s_{l-1}^{m*}c_l^m\right),
\label{Yz}
\end{eqnarray}
where $a_l^m$ and $s_l^m$ are the beam-shape 
and the scattering coefficients,
the symbol $^*$ means complex conjugation, $c_l^m = \sqrt{(l-m)(l+m)/[(2l-1)(2l+1)]}$,
`Re' and `Im' denote the real and the imaginary parts,
and
\begin{eqnarray}
\nonumber
 Y_\perp &=& \frac{i}{2\pi (ka)^2} \sum_{l,m}\biggl[
(a_l^{m*}+s_l^{m*}) 
( s_{l+1}^{m-1}b_{l+1}^{-m} + s_{l-1}^{m-1}b_l^{m-1}) \\
  &+& (a_l^{m}+s_l^{m})(s_{l+1}^{m+1*}b_{l+1}^m + s_{l-1}^{m+1*}b_l^{-m-1})\biggr],
\label{Yp}
\end{eqnarray}
with $b_l^m = \sqrt{(l+m)(l+m+1)/[(2l-1)(2l+1)]}$.

The beam-shape coefficients are determined from the incident pressure amplitude as follows~\cite{silva:298,mitri392}
\begin{equation}
 a_l^m = \frac{1}{p_0 j_l(k b)}\int_{4\pi} \hat{p}_i(k b,\theta,\varphi)  
Y_l^{m*}(\theta,\varphi)  d\Omega,
\label{alm}
\end{equation}
where $p_0$ is the incident pressure magnitude, $j_l$ is the spherical Bessel function of 
$l$-th order and $d\Omega = \sin \theta d\theta d\varphi$.
The integral is performed over the surface
of a control spherical region of radius $b$. 
This region encloses the inclusion that is subjected to the radiation force.
The integral in Eq.~(\ref{alm}) is the spherical harmonic transform (SHT) of the incident pressure amplitude.
For an arbitrary pressure amplitude the integral in equation~(\ref{alm}) should be evaluated numerically.

The scattering coefficient for an absorbing fluid sphere can obtained from the following boundary conditions:
the pressure and the radial component of the particle velocity should be continuous
across the sphere's surface. 
Accordingly, we have~\cite{silva:epl}
\begin{equation}
s_l^m = a_l^m \det
\left[
\begin{array}{cc}
 \gamma j_l(k a) & j_l(k_1 a)\\
j_l'(k a) & j_l'(k_1 a)
\end{array}
\right] \det\mbox{}^{-1}
\left[
\begin{array}{cc}
 -\gamma h_l^{(1)}(k a) & j_l(k_1 a)\\
-{h_l^{(1)}}'(k a) & j_l'(k_1 a)
\end{array}
\right]  
\label{slm}
\end{equation}
where $\gamma= k_1 \rho_0 /k \rho_1 $ is the impedance index, $k_1=\omega/c_1 + i\alpha$, with
 $\alpha$ being the absorption coefficient,
$h_l^{(1)}$ is the first-kind spherical Hankel of $l$-th order, and the prime symbol means differentiation.

\section{Acoustic Bessel beam}
The pressure amplitude of a $n$-th order Bessel beam propagating along the $z$-axis is 
expressed as~\cite{mcgloin:15}
\begin{equation}
\label{bessel}
 \hat{p}_{i,n} =   e^{i k z\cos \beta } J_n(k \varrho \sin \beta )e^{i n\varphi},
\end{equation}
where $J_n$ is the $n$th-order Bessel function, $\beta$ 
is the half-cone angle of the beam,
$\varrho = \sqrt{(x-x_0)^2 + (y-y_0)^2}$, and $\varphi = \tan^{-1} [(y-y_0)/(x-x_0)]$.
The relative position of the beam axis with respect to the center of the sphere
in the $xy$-plane is denoted by $(x_0,y_0)$.

The beam-shape coefficients for an
off-axial Bessel beam need to be numerically computed from the SHT given in equation~(\ref{alm}).
This is performed through the discrete SHT (DSHT), which is realized through the fast Fourier transform (FFT) for the integration in the polar angle and 
the Gauss-Legendre quadrature used for the azimuthal angle integration~\cite{colton:p79}.
In this algorithm, the pressure amplitude is sampled on the control sphere surface.
In the azimuthal direction, this function has $M$ uniformly distributed sampling points,
while in the polar direction, it has $N$ points corresponding to the zeros of the Legendre polynomial of $N$-th order.

\section{Results and discussion}
The radiation force caused by a zero- and a first-order Bessel beams upon a sphere suspended 
in water ($\rho_0=\unit[1000]{kg/m^3}$ and $c_0=\unit[1500]{m/s}$) is analyzed here.
The sphere is made of silicone oil~\cite{schroder:565} 
($\rho_1= \unit[970]{kg/m^3}$, $ c_1= \unit[1004]{m/s}$, and $\alpha=\unit[2 \times 10^{-10} f^{1.7}]{Np\cdot Hz^{-1.7}/m}$, with $f$ being the frequency).
The silicone oil was chosen because it may form drops into water without mixing.
The frequency of the incident beams is fixed to $\unit[1]{MHz}$.

To compute the radiation force, we have to obtain the beam-shape $a_l^m$ and then
the scattering coefficients given in equation~(\ref{slm}).
To compute the beam-shape coefficient we use the DSHT algorithm with $M=2^{12}$ and $N=256$.
In this case, the mean absolute error between the pressure amplitude of the Bessel beams given in 
equation~(\ref{bessel}) and reconstructed waves using the beam-shape coefficients is about $10^{-9}$.
Furthermore, the infinite series in equations~(\ref{Yz}) and (\ref{Yp}) should be truncated at $l=L$.
The truncation order $L$ should be the smallest integer
for what the scattering coefficient ratio $|s_L^m/s_0^0|\ll 1$.
This is required to ensure proper convergence of the radiation force series. 
In the performed computations, the truncation order is such that the scattering coefficient ratio is as small as $10^{-9}$.

To analyze the behavior of the axial radiation force, 2D plots of the radiation force $Y_z$ are computed in the range
$ 0^\circ \le \beta \le 90^\circ$ and $0\le ka \le 10$. 
As an initial test, we have recovered previous results 
based on the analytical calculation of the
axial radiation force upon an hexane sphere in 
the on-axis configuration~\cite{mitri245202}.
For the sake of brevity, these results will not be shown here.

\begin{figure}[h!]
\centering
\subfigure[]{
\includegraphics[scale=.5]{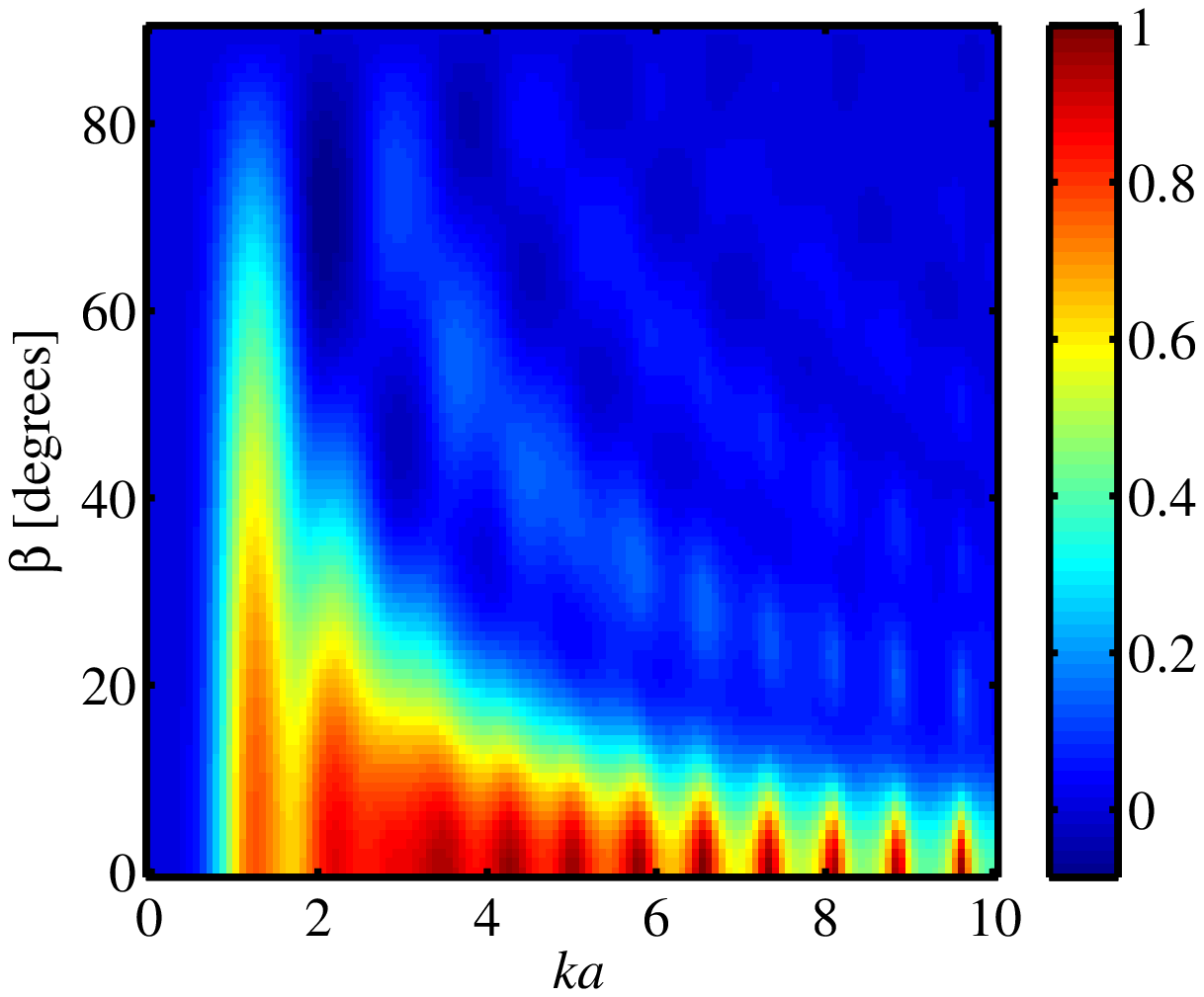}
\label{fig:bb0_beta_ka_a}}
\subfigure[]{
\includegraphics[scale=.5]{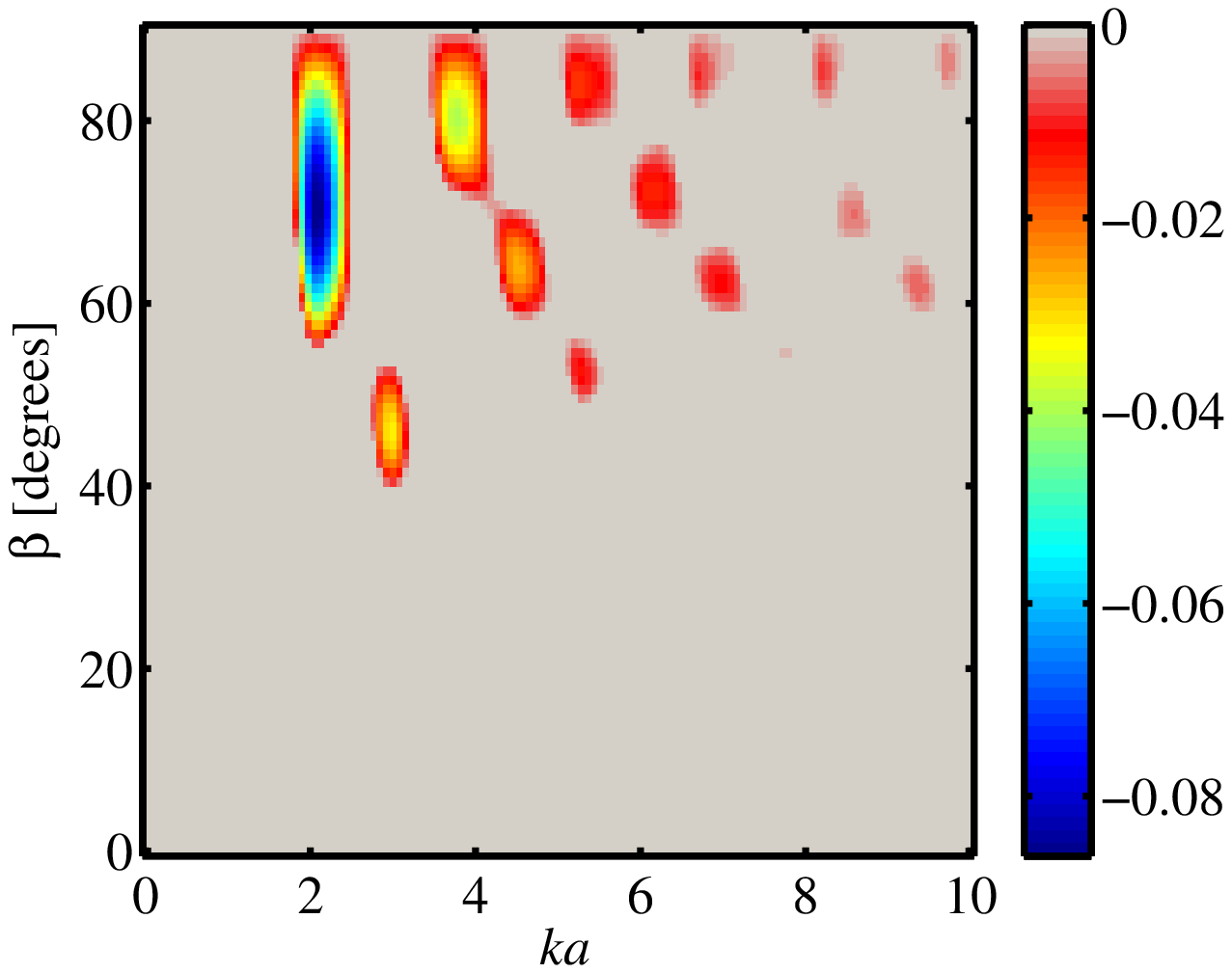}
\label{fig:bb0_beta_ka_b}}
\subfigure[]{
\includegraphics[scale=.5]{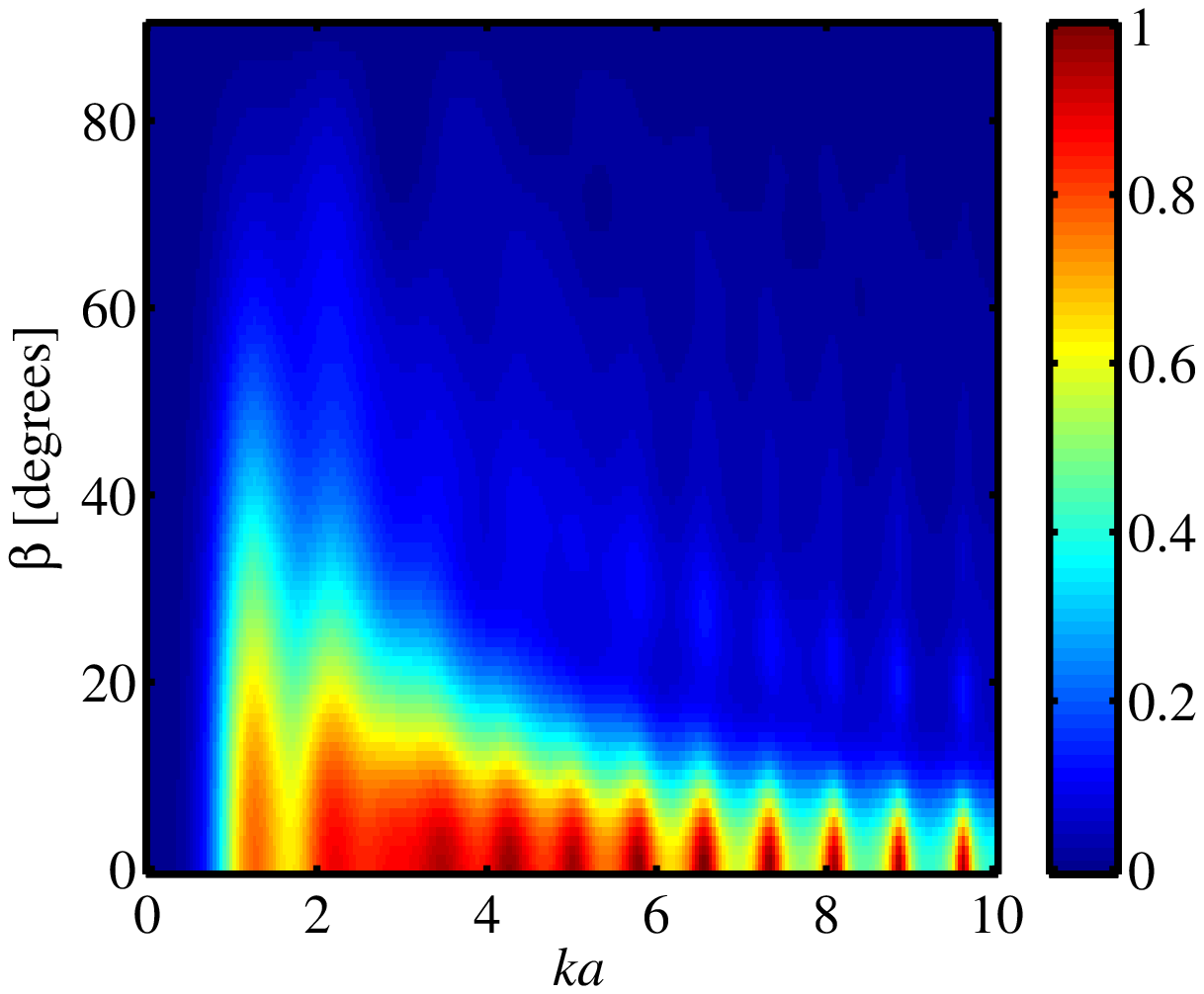}
\label{fig:bb0_beta_ka_c}}
\subfigure[]{
\includegraphics[scale=.5]{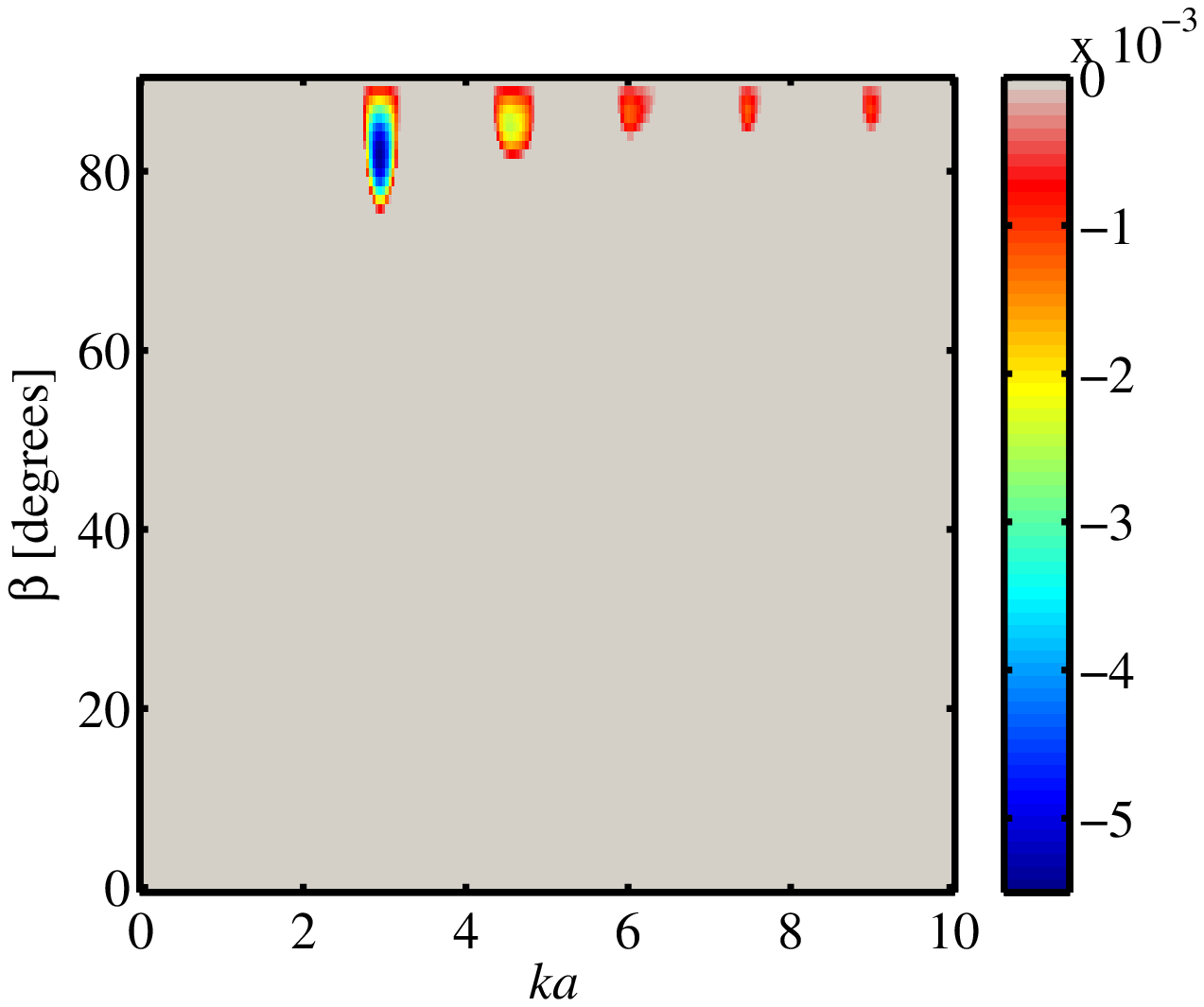}
\label{fig:bb0_beta_ka_d}}
\caption{Axial radiation force function $Y_z$ for the silicone oil sphere due to the zero-order Bessel beam.
The function $Y_z$ is normalized to $1.3931$.
(a) The sphere is placed on the axis of the beam.
(b) The islands of  attractive force in the on-axis configuration.
(c) The sphere location has a offset of $kx_0=1.6$ along the $x$-axis. 
(d) The islands of attractive force in the off-axial configuration.
\label{fig:bb0_beta_ka}}
\end{figure}

The 2D plots of the axial radiation force function $Y_z$ produced
by the zero-order Bessel beam 
are depicted in figure~\ref{fig:bb0_beta_ka}.
The radiation force function is normalized to
$1.3931$.
The on-axis configuration is shown in figures~\ref{fig:bb0_beta_ka_a} and \ref{fig:bb0_beta_ka_b}.
In this configuration, islands of negative radiation force (tractor beam) are clearly spotted.
The largest magnitude of the negative axial force lies
in the region around $ka=2$ and $60^\circ<\beta<80^\circ$.
In this region, the attractive force magnitude corresponds to
approximately $8\%$ of the maximum value of the repulsive axial force.
With an offset of $kx_0=1.6$ along the $x$-axis, 
the islands of attractive force shift toward higher $ka$
values.
Moreover, the strength of the negative force drops one order of magnitude.

\begin{figure}[h]
\centering
\subfigure[]{
\includegraphics[scale=.5]{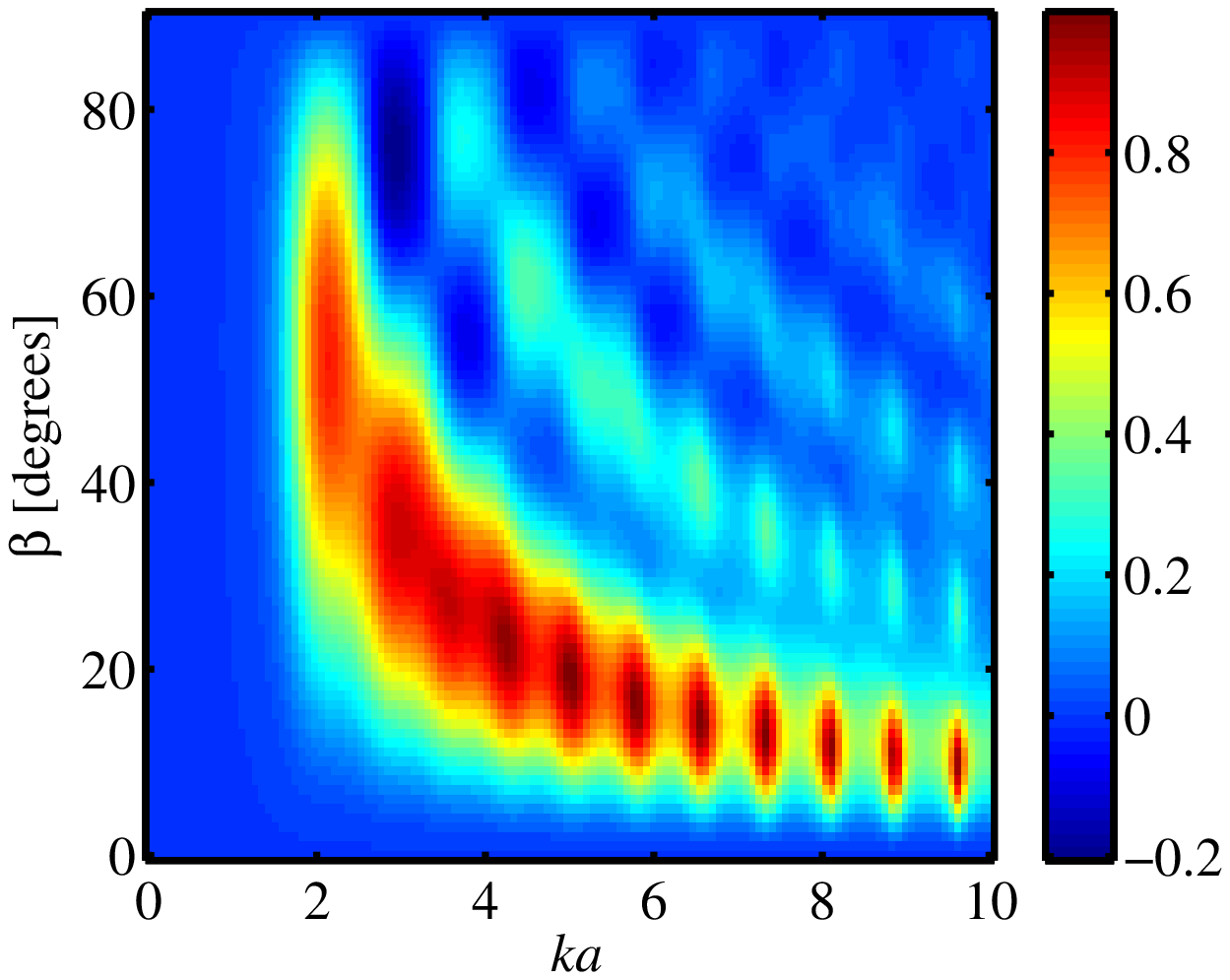}
\label{fig:bb1_beta_ka_a}}
\subfigure[]{
\includegraphics[scale=.5]{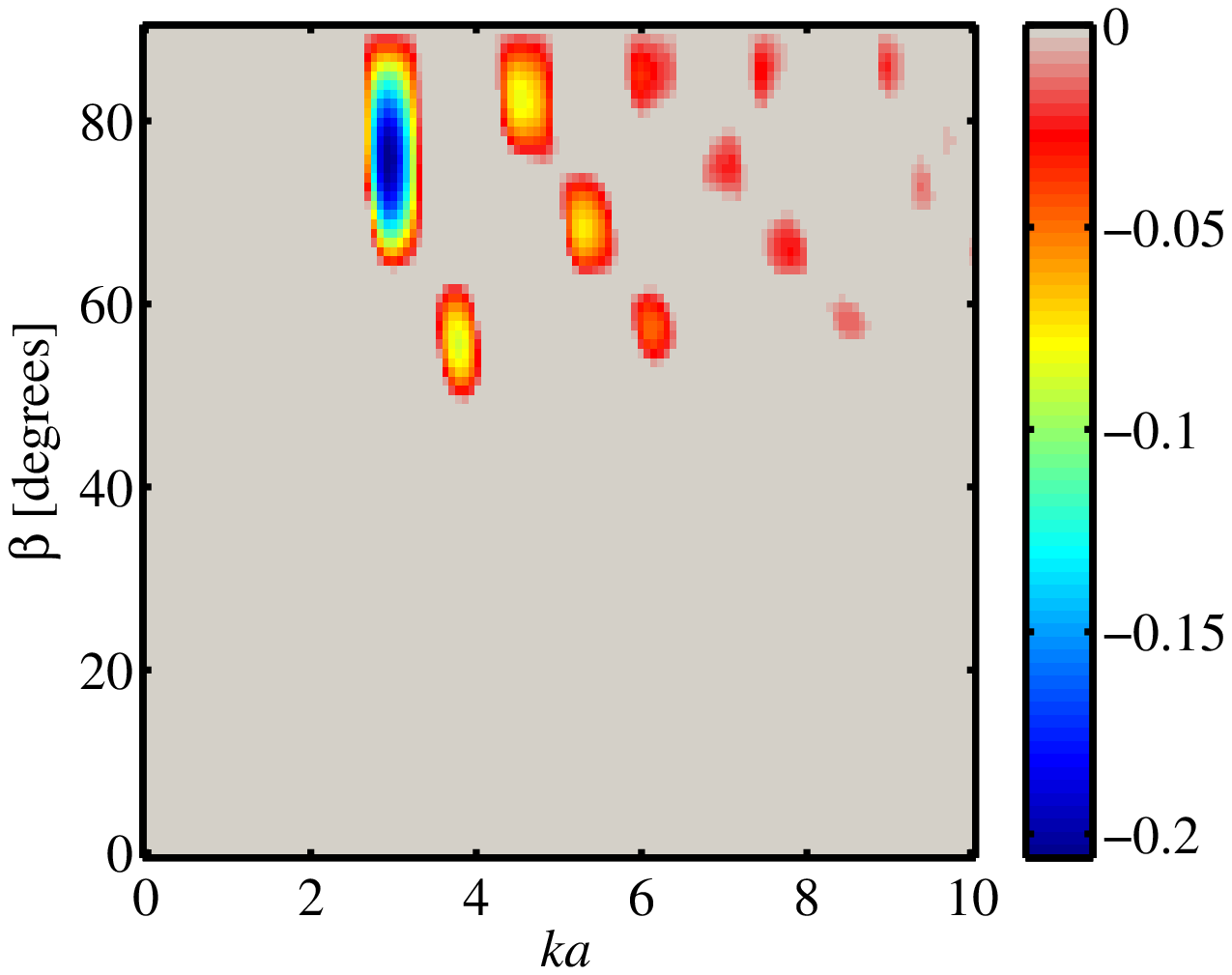}
\label{fig:bb1_beta_ka_b}}
\subfigure[]{
\includegraphics[scale=.5]{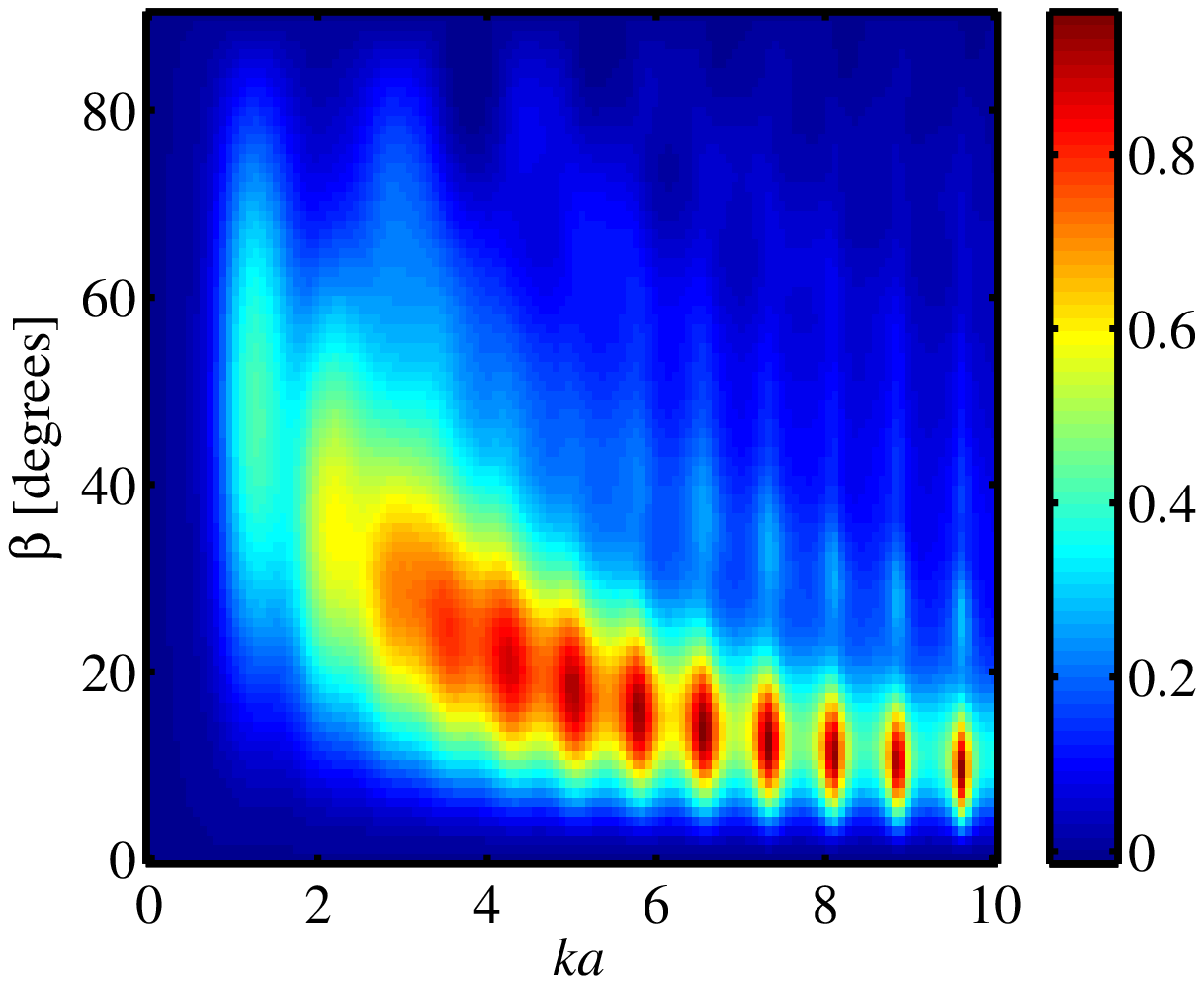}
\label{fig:bb1_beta_ka_c}}
\subfigure[]{
\includegraphics[scale=.5]{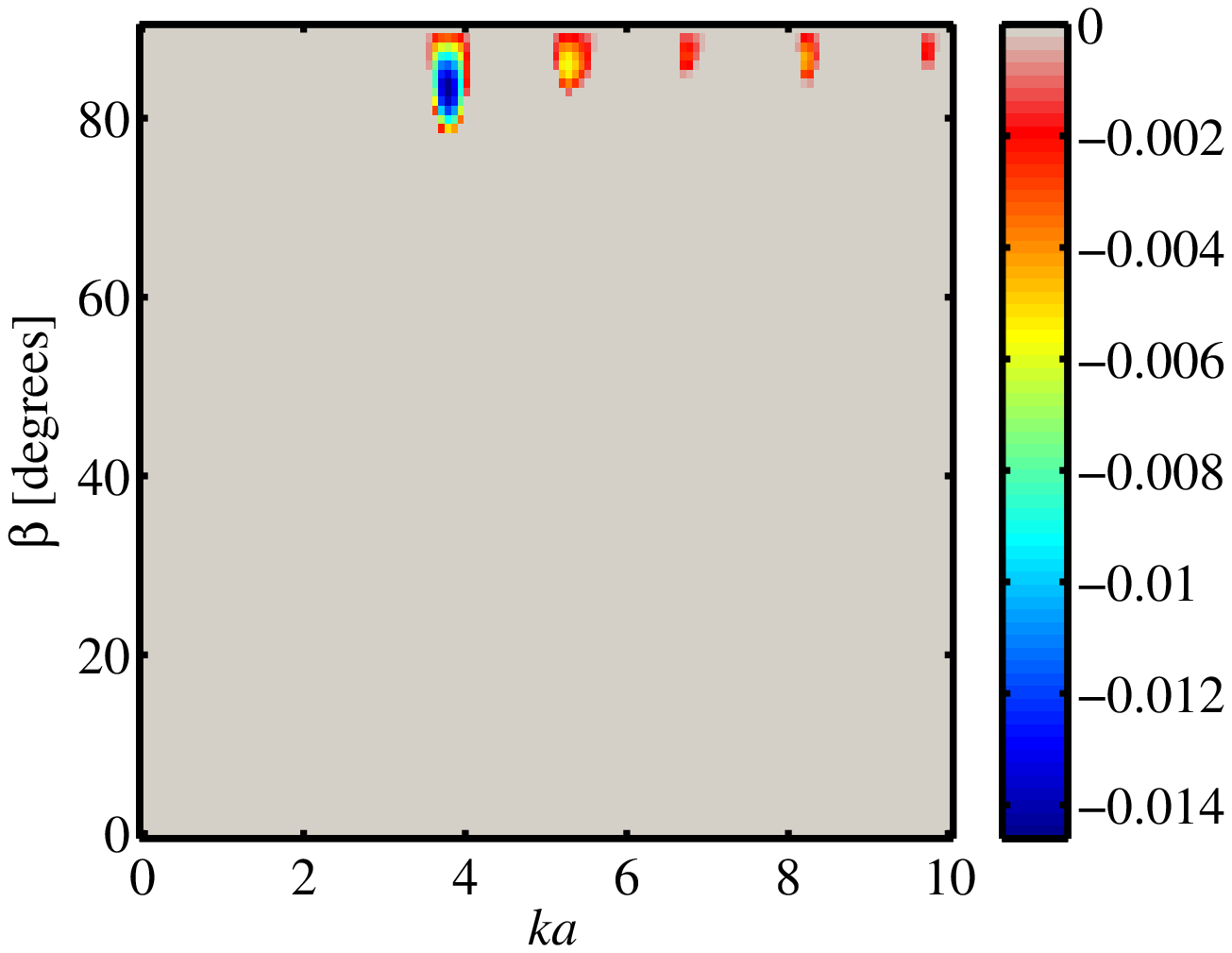}
\label{fig:bb1_beta_ka_d}}
\caption{Axial radiation force function $Y_z$ for the silicon oil sphere due to the first-order Bessel beam.
The function $Y_z$ is normalized to $0.4201$.
(a) The sphere is placed on the beam's axis.
(b) The islands of attractive radiation force in the on-axis configuration.
(c) The sphere location has a offset of $kx_0=1.6$ along the $x$-axis. 
(d) The islands of attractive radiation force in the off-axial configuration.\label{fig:bb1_beta_ka}}
\end{figure}

Results of the axial radiation force function due to the first-order Bessel vortex beam as a function
of $\beta$ and $ka$ are shown in figure~\ref{fig:bb1_beta_ka}.
The radiation force function is normalized to $0.4201$.
In the on-axis configuration (figures~\ref{fig:bb1_beta_ka_a} and \ref{fig:bb1_beta_ka_b}), 
islands of attractive radiation force arise.
The largest magnitude of the negative force takes place within the region $65^\circ<\beta<80^\circ$ and $ka=3$.
The attractive force here is about $20\%$ lower than
that due to the zero-order Bessel beam.
When an offset of $kx_0=1.6$ is applied to beam along the  $x$-axis, the islands of attractive force move toward
higher $ka$ values.
Furthermore, with this offset the attractive force islands shrink and their strength drop by one order of magnitude.

\begin{figure}[t]
\centering
\subfigure[]{
\includegraphics[scale=.5]{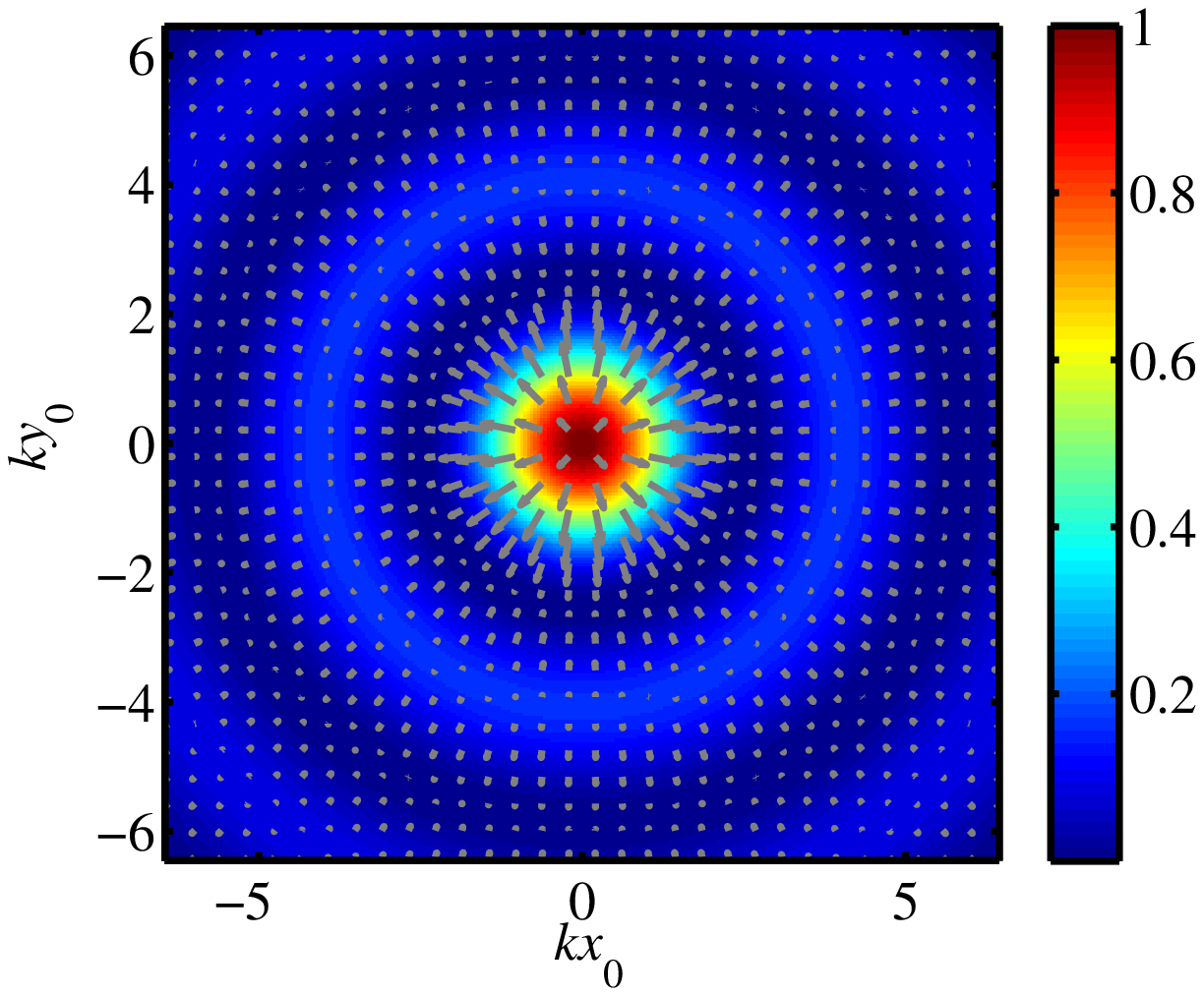}
\label{fig:bb0_quiver_a}}
\subfigure[]{
\includegraphics[scale=.5]{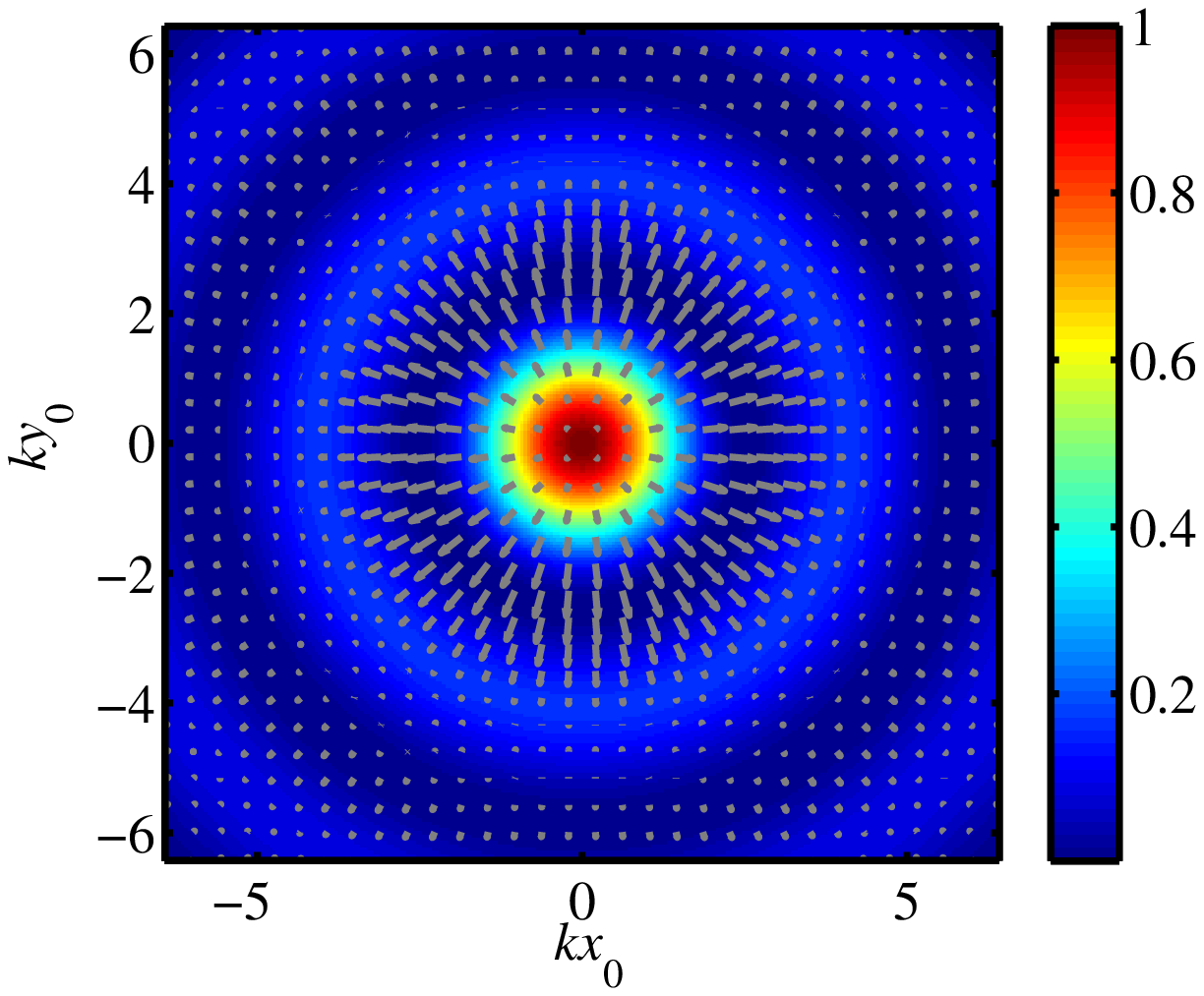}
\label{fig:bb0_quiver_b}}
\caption{Transverse radiation force field ${\bf Y}_\perp=(Y_x,Y_y)$ caused by the zero-order 
Bessel beam with $\beta=70^\circ$ for the silicone sphere with (a) $ka=0.1$
and (b) $ka=2$.
The background images corresponds to the beam intensity.
\label{fig:bb0_quiver}}
\end{figure}

The transverse radiation force field ${\bf Y}_\perp=(Y_x,Y_y)$ caused by the zero-order Bessel beam with $\beta=70^\circ$ 
for the silicone sphere  is illustrated in figure~\ref{fig:bb0_quiver}.
The transverse force fields are computed by varying the relative position between the sphere and the beam 
(offset) within $-6.4\le kx_0,ky_0\le 6.4$.
Furthermore, the corresponding vector fields are plotted on top of the time-averaged energy flux (intensity) of
the incident beam normalized to the unit.
Two different sphere's size factors are considered, which correspond to a repulsive $(ka=0.1)$ and an attractive 
$(ka=2)$ axial radiation force (see figure~\ref{fig:bb0_beta_ka}).
In both cases, the central spot of the beam is a region of unstable equilibrium for the sphere.
The lines of force diverge from the beam axis. 

\begin{figure}
\centering
\subfigure[]{
\includegraphics[scale=.45]{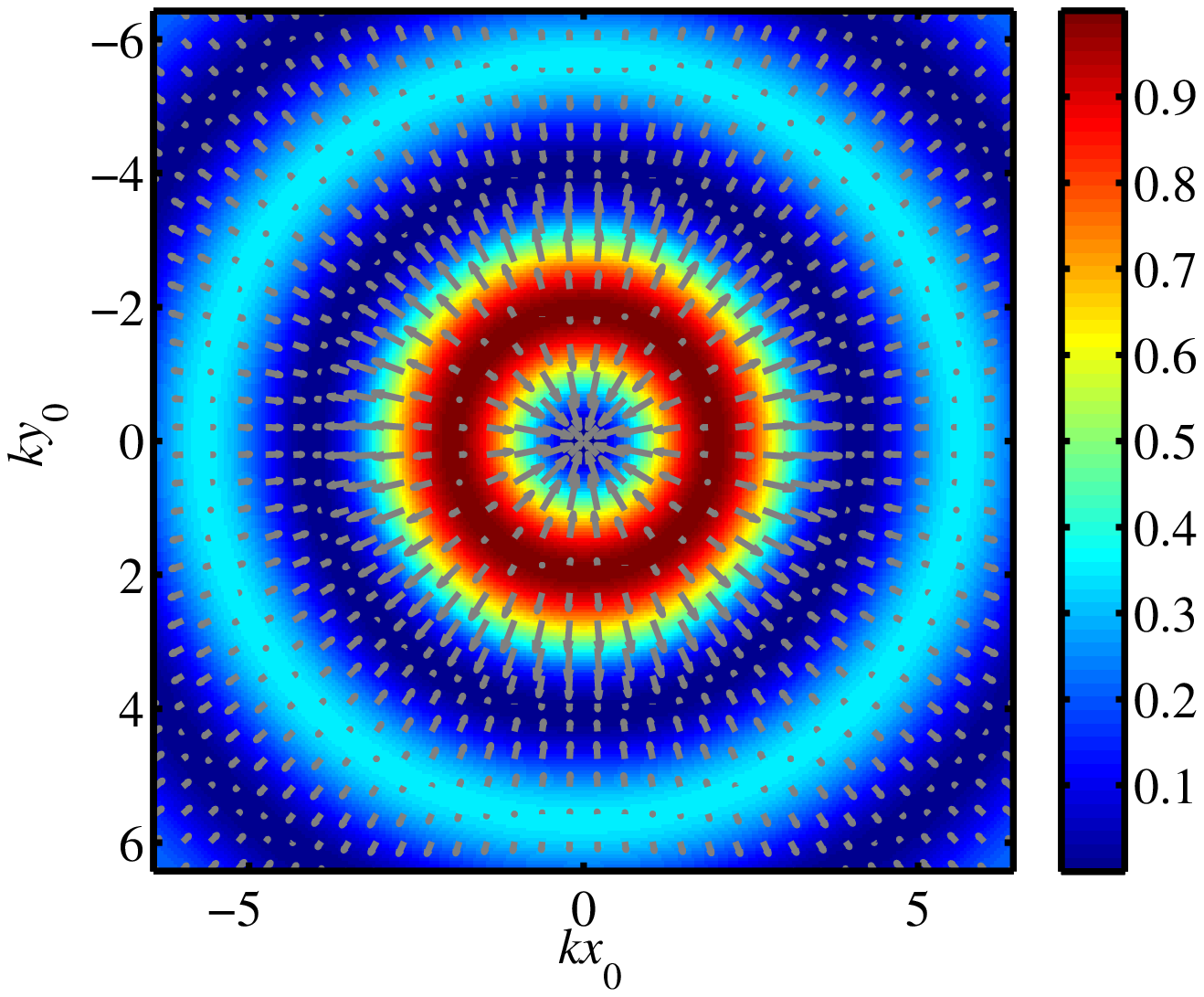}
\label{fig:bb1_quiver_a}}
\subfigure[]{
\includegraphics[scale=.45]{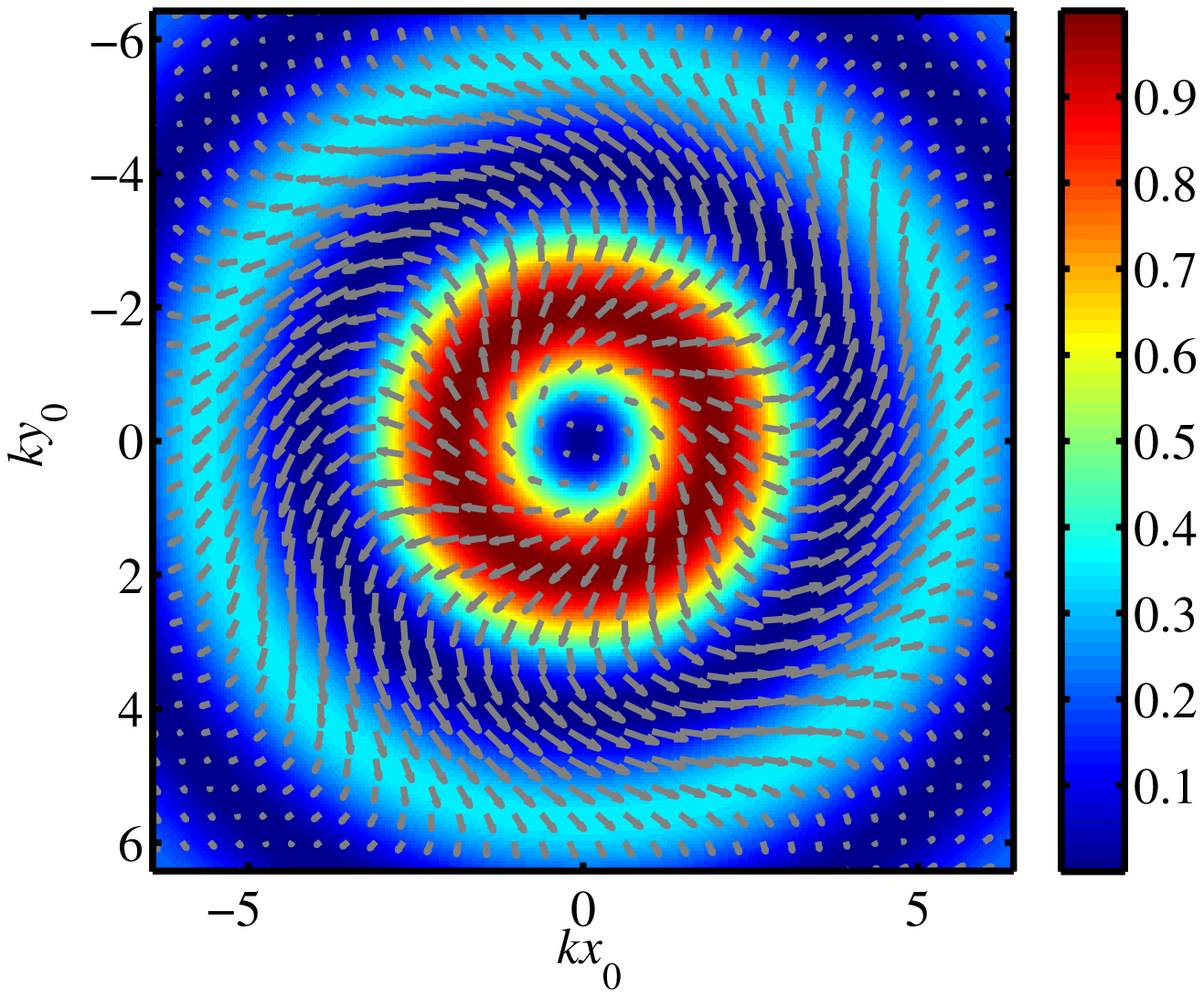}
\label{fig:bb1_quiver_b}}
\caption{Transverse radiation force field ${\bf Y}_\perp=(Y_x,Y_y)$ caused by 
the first-order Bessel beam with $\beta=70^\circ$ upon the silicone sphere with (a) $ka=0.1$
and (b) $ka=3$.
The background images correspond to the beam intensity.
\label{fig:bb1_quiver}}
\end{figure}

In figure~\ref{fig:bb1_quiver}, the transverse radiation force field due to the first-order Bessel beam with $\beta=70^\circ$
for the silicone sphere is depicted.
The force fields are computed similarly to those in figure~\ref{fig:bb0_quiver}.
To analyze the pressor and the tractor beam two size factors are chosen, namely $ka=0.1$ and $ka=3$.
For $ka=0.1$ the beam central spot is a region of stable equilibrium for the sphere.
The sphere is pushed by the pressor beam along the $z$-axis.
The transverse force points inwardly toward the beam axis.
Thus, the sphere can be trapped in the beam's axis.
As the sphere is moved radially away from the beam's axis, the transverse radiation force
reverses its direction.
It flips from inward to outward direction as the spheres varies from the first maximum 
to the first minimum of the beam's intensity.
Moreover, the transverse radiation force points inwardly in the region from the first minimum to the second maximum 
beam's intensity.
In this region, the sphere might be trapped.
The transverse radiation force field changes considerably when $ka=3$.
It points outwardly following the clockwise direction as the sphere moves radially away from the beam's axis.
Moreover, inside the beam's central spot, the sphere is pulled back along the $z$-axis by the tractor beam.
Moving the sphere radially from the first maximum to the first minimum, the transverse radiation force
points outwardly along the counterclockwise direction.
No stable equilibrium region is formed in this case.

\section{Conclusion}
The radiation force produced by a zero- and a first-order Bessel beam upon a silicone-oil sphere in
both on- and off-axial configurations has been studied.
The analysis was based on the 3D radiation force expressions given in terms of the beam-shape and
the scattering coefficients.
The beam-shape coefficients were numerically computed through the DSHT algorithm, while the scattering 
coefficients were obtained by applying
the acoustic boundary conditions across the sphere's surface.
With appropriate selection of the sphere size factor $ka$ and the beam half-cone angle $\beta$,
the axial radiation force exerted on the sphere might be repulsive (pressor beam) or attractive (tractor beam).
As the sphere departs radially from the beam's axis, the attractive axial radiation force becomes weaker.
In addition, the transverse radiation force field was computed showing that the silicone sphere can be trapped by 
using the first-order Bessel beam.
These results represent an important step toward the design of a new generation of acoustical tweezers 
operating with Bessel beams for potential applications in bioengineering, biophysics, and other related fields.

\section*{Acknowledgments}
This work was supported by grants 306697/2010-6 and 477653/2010-3 CNPq (Brazilian agency).

\section*{References}


\end{document}